\documentclass[twocolumn,aps,prl,showpacs,superscriptaddress,amsfonts,amssymb,amsmath]{revtex4}
\usepackage{graphicx}
\usepackage{amsmath}
\allowdisplaybreaks[4]
\usepackage{color,xcolor}

\def\a{\alpha}
\def\b{\beta}

\def\V={{{\bf\rm{V}}}}

\def\beq{\begin{equation}}
\def\eeq{\end{equation}}
\def\bea{\begin{eqnarray}}
\def\eea{\end{eqnarray}}
\def\ba{\begin{array}}
\def\ea{\end{array}}
\def\no{\nonumber}

\def\lt{\left}
\def\rt{\right}

\begin{document}

\title{Exact surface energy and helical spinons in the XXZ spin chain with arbitrary non-diagonal boundary fields}
\author{Yi Qiao}
\affiliation{Beijing National Laboratory for Condensed Matter
Physics, Institute of Physics, Chinese Academy of Sciences, Beijing
100190, China}
\author{Junpeng Cao}
\affiliation{Beijing National Laboratory for Condensed Matter
Physics, Institute of Physics, Chinese Academy of Sciences, Beijing
100190, China}
\affiliation{School of Physical Sciences, University of Chinese Academy of
Sciences, Beijing, China}
\affiliation{Songshan Lake Materials Laboratory, Dongguan, Guangdong 523808, China}
\affiliation{Peng Huanwu Center for Fundamental Theory, Xian 710127, China}
\author{Wen-Li Yang}
\thanks{wlyang@nwu.edu.cn}
\affiliation{Peng Huanwu Center for Fundamental Theory, Xian 710127, China}
\affiliation{Institute of Modern Physics, Northwest University, Xian 710127, China}
\affiliation{School of Physics, Northwest University, Xian 710127, China}
\affiliation{Shaanxi Key Laboratory for Theoretical Physics Frontiers, Xian 710127, China}
\author{Kangjie Shi}
\affiliation{Institute of Modern Physics, Northwest University, Xian 710127, China}
\author{Yupeng Wang}
\thanks{yupeng@iphy.ac.cn}
\affiliation{Beijing National Laboratory for Condensed Matter
Physics, Institute of Physics, Chinese Academy of Sciences, Beijing
100190, China}
\affiliation{Peng Huanwu Center for Fundamental Theory, Xian 710127, China}
\affiliation{The Yangtze River Delta Physics Research Center, Liyang, Jiangsu 213300, China}

\begin{abstract}
An analytic method is proposed to compute the  surface energy and elementary excitations of the XXZ spin chain with generic non-diagonal boundary fields. For the gapped case,  in some boundary parameter regimes the contributions of the two boundary fields to the surface energy are non-additive. Such a correlation effect between the two boundaries also depends on the parity of the site number $N$ even in the thermodynamic limit $N\to\infty$. For the gapless case, contributions of the two boundary fields to the surface energy are additive due to the absence of long-range correlation in the bulk. Although the $U(1)$ symmetry of the system is broken, exact spinon-like excitations, which obviously do not carry spin-$\frac12$, are observed. The present method provides an universal procedure to deal with quantum integrable systems either with or without $U(1)$ symmetry.
\end{abstract}

\pacs{75.10.Pq, 03.65.Vf, 71.10.Pm}

\maketitle

Quantum integrable systems with generic non-diagonal boundary fields have attracted a lot of attentions since their important applications in high energy physics \cite{Ber05}, open string/gauge theory \cite{Sch06,Bei12,Che18}, condense matter physics \cite{And20} and non-equilibrium statistical physics \cite{Jde05, Sir09}.
However, how to compute the physical quantities of such kind of systems has puzzled people for quite a long time.
In the past several decades, many efforts have been made to approach this tough problem \cite{alcaraz,Klu92, Ess05, skorik,cao03,nep02, cysw, book, liyuanyuan, Bas13, BP}
but only under some special conditions the physical quantities can be calculated. Formally, the exact spectra of quantum integrable models without $U(1)$ symmetry can be expressed in terms of inhomogeneous $T-Q$ relations \cite{cysw,book}.
Even though, to study their physical properties based on the inhomogeneous Bethe ansatz equations is still quite hard
because of the complicated patterns of Bethe roots in the complex plane.

In this Letter, we propose a novel analytic method to study the surface energy and elementary excitations of the XXZ spin chain with arbitrary non-diagonal boundary fields. Our central idea lies in that instead of the Bethe roots, we use the zero roots of the transfer matrix to parameterize the spectrum. Starting from a transfer matrix including proper site-dependent inhomogeneity (described by a density $\sigma(\theta)$ in the thermodynamic limit), the density of the zero roots of the homogeneous transfer matrix, which is crucial to compute the physical quantities, can be obtained via analytic continuation.

The model Hamiltonian we shall consider reads
\begin{eqnarray}
&&H =\sum_{j=1}^{N-1}\lt\{\sigma_j^x\sigma_{j+1}^x+\sigma_j^y\sigma^y_{j+1}+\cosh\eta\sigma_j^z\sigma_{j+1}^z\rt\}\no\\
&& \qquad +{\vec h}_-\cdot{\vec \sigma}_1+{\vec h}_+\cdot{\vec \sigma}_N,\label{Hamiltonian}
\end{eqnarray}
where $\sigma_{j}^{\alpha}$ $(\alpha=x, y, z)$ are the Pauli matrices on site $j$, $\eta$ is the anisotropic parameter and $\vec{h}_\pm\equiv (h_\pm^x, h_\pm^y, h_\pm^z)$ are the boundary fields
\begin{eqnarray}
&&h^z_\pm=\mp\frac{\sinh\eta\cosh\alpha_\pm\sinh\beta_\pm}{\sinh\alpha_\pm\cosh\beta_\pm},\no\\
&&h^x_\pm=\frac{\sinh\eta\cos\theta_\pm}{\sinh\alpha_\pm\cosh\beta_\pm},\;\;
h^y_\pm=\frac{\sinh\eta\sin\theta_\pm}{\sinh\alpha_\pm\cosh\beta_\pm}, \label{H-b}
\end{eqnarray}
characterized by the boundary parameters $\alpha_{\pm}$, $\beta_{\pm}$ and $\theta_{\pm}$.
The Hamiltonian (\ref{Hamiltonian}) is generated by the transfer matrix $t(u)$ as
\begin{eqnarray}
H=\sinh\eta \frac{\partial \ln t(u)}{\partial
u}|_{u=0,\{\theta_j=0\}}-c_0,\label{Ham-1}
\end{eqnarray}
where $\{\theta_j|j=1,\cdots,N\}$ are the inhomogeneity parameters, $c_0=N\cosh\eta +\tanh\eta\sinh\eta$,  $t(u)$ is defined as \cite{sklyanin88}
\bea
&&t(u)=tr_0\{K^+_0(u)R_{0N}(u-\theta_N)\cdots
R_{01}(u-\theta_1)\nonumber \\
&&\qquad \quad\times K^-_0(u)R_{10}(u+\theta_1)\cdots
R_{N0}(u+\theta_N)\}.\label{trans}\eea
Here $K^-_0(u)$ is the boundary reflection matrix on one end of the spin chain
\bea
&&K^-(u)=\lt(\begin{array}{ll}K^-_{11}(u)&K^-_{12}(u)\\
K^-_{21}(u)&K^-_{22}(u)\end{array}\rt),\no\\
&&K^-_{11}(u)=2\sinh\a_-\cosh\b_{-}\cosh u\no\\
&&\qquad\qquad +2\cosh\a_-\sinh\b_-\sinh u,\no\\
&&K^-_{12}(u)=e^{-i\theta_-}\sinh(2u),\;\;
K^-_{21}(u)=e^{i\theta_-}\sinh(2u),\no \\
&&K^-_{22}(u)=2\sinh\a_-\cosh\b_{-}\cosh u\no\\
&&\qquad\qquad-2\cosh\a_-\sinh\b_-\sinh u,
\label{K-matrix}\eea
and $K^+_0(u)$ is the dual boundary  matrix on the other end
\bea
K^+(u)=\lt.K^-(-u-\eta)\rt|_{(\a_-,\b_-,\theta_-)\rightarrow
(-\a_+,-\b_+,\theta_+)}.\label{K-6-2} \eea
The six-vertex $R$-matrix
\bea
&&R_{0,j}(u)=\frac{\sinh(u+\eta)+\sinh u}{2\sinh \eta}+\frac{1}{2} (\sigma^x_j \sigma^x_0 +\sigma^y_j \sigma^y_0) \nonumber\\
 &&\qquad\qquad+ \frac{\sinh(u+\eta)-\sinh u}{2\sinh \eta} \sigma^z_j \sigma^z_0,
\label{r-matrix} \eea
satisfies the Yang-Baxter equation (YBE) \cite{yang2,bax1}
and the reflection matrices satisfy the reflection equation (RE) or the dual one \cite{sklyanin88,cherednik,devega,zama}.
The YBE and REs lead to that the transfer matrices with different spectral parameters commute mutually, i.e., $[t(u),t(v)]=0$,
which ensures the integrability of the model (\ref{Hamiltonian}).

Given an arbitrary eigenvalue $\Lambda(u)$ of the transfer matrix $t(u)$, we have the identities \cite{npb14}
\begin{eqnarray}
&&\Lambda(\theta_j)\Lambda(\theta_j-\eta)=a(\theta_j)a(-\theta_j), \quad j=1,\cdots, N,\label{tw1}\\
&&\Lambda(0)=a(0),\quad \Lambda\Big(\frac{i\pi}2\Big)=a\Big(\frac{i\pi}2\Big),\label{tw2}
\end{eqnarray}
with
\bea
&&a(u)=-4
\frac{\sinh(2u+2\eta)}{\sinh(2u+\eta)}
\sinh(u-\a_-)\cosh(u-\b_-)\no\\
&&\qquad\quad \times
\sinh(u-\a_+)
\cosh(u-\b_+)\nonumber \\
&&\qquad\quad \times\prod_{l=1}^N\frac{\sinh(u-\theta_l+\eta)\sinh(u+\theta_l+\eta)}
{\sinh^2\eta}.
\eea
From the definition of $t(u)$ in (\ref{trans}), we deduce that $\Lambda(u)$ is a degree $2N+4$ trigonometric polynomial of $u$. It also possesses the properties
$\Lambda(u)=\Lambda(-u-\eta)$ and $\Lambda(u+i\pi)=\Lambda(u)$.
Thus we can parameterize the eigenvalue $\Lambda(u)$ by its roots $\{z_j\}$ as
\bea
\Lambda(u)=\Lambda_0\prod_{j=1}^{N+2}\sinh(u-z_j+\frac{\eta}{2})
\sinh(u+z_j+\frac{\eta}{2}). \label{Expansion-L}
\eea
$\Lambda_0=-8\cos(\theta_--\theta_+)\sinh^{-2N}\eta$ is determined by the asymptotic behavior of $t(u)$ when $u\to\infty$. In such a sense, Eqs.(\ref{tw1}-10) determine the roots $\{z_j|j=1,\cdots,N+2\}$ completely for a given set
of inhomogeneity parameters. In the homogeneous limit $\{\theta_j=0|j=1,\cdots, N\}$, Eq.(\ref{tw1}) is replaced by \cite{cysw}
\bea
[\Lambda(u)\Lambda(u-\eta)]^{(n)}|_{u=0}=[a(u)a(-u)]^{(n)}|_{u=0}, \label{H-BAE1}
\eea
where the superscript $(n)$ indicates the $n$-th order derivative and $n=0,1,\cdots, N-1$.
From (\ref{Ham-1}) and (\ref{Expansion-L}), the eigenvalues of the Hamiltonian (\ref{Hamiltonian}) can be expressed  as
\bea
E=\sinh\eta\sum_{j=1}^{N+2}[\coth(z_j+\frac{\eta}{2})-\coth (z_j-\frac{\eta}{2})]-c_0.\label{Energy}
\eea

\begin{figure}[ht]
\centering
\includegraphics[width=4.2cm,height=3.2cm]{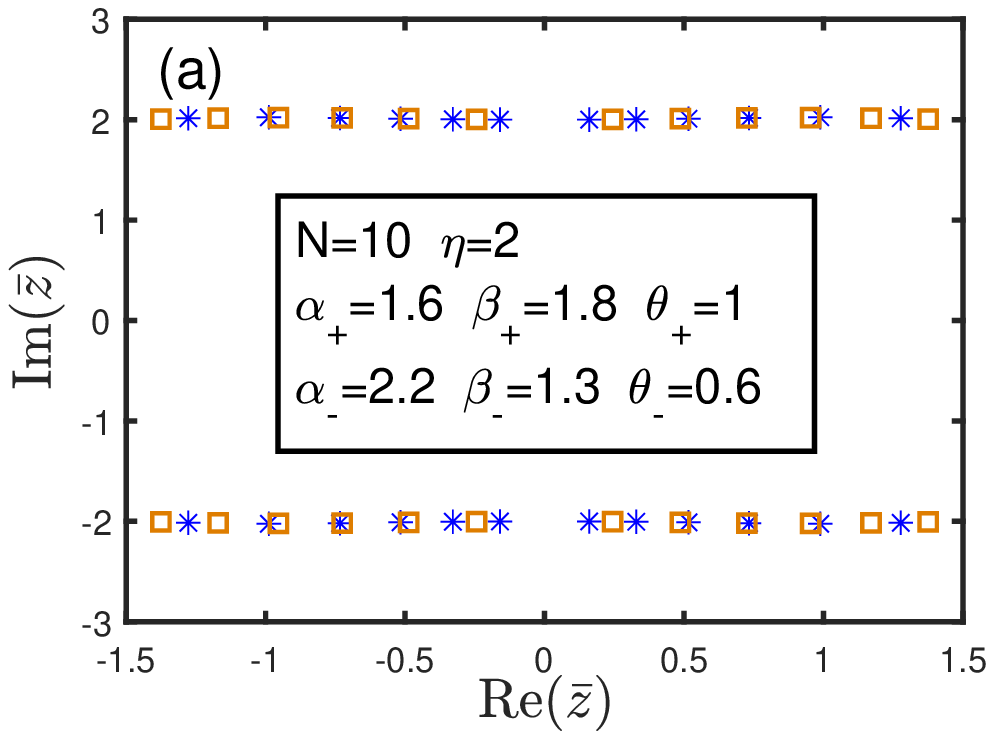}
\includegraphics[width=4.2cm,height=3.2cm]{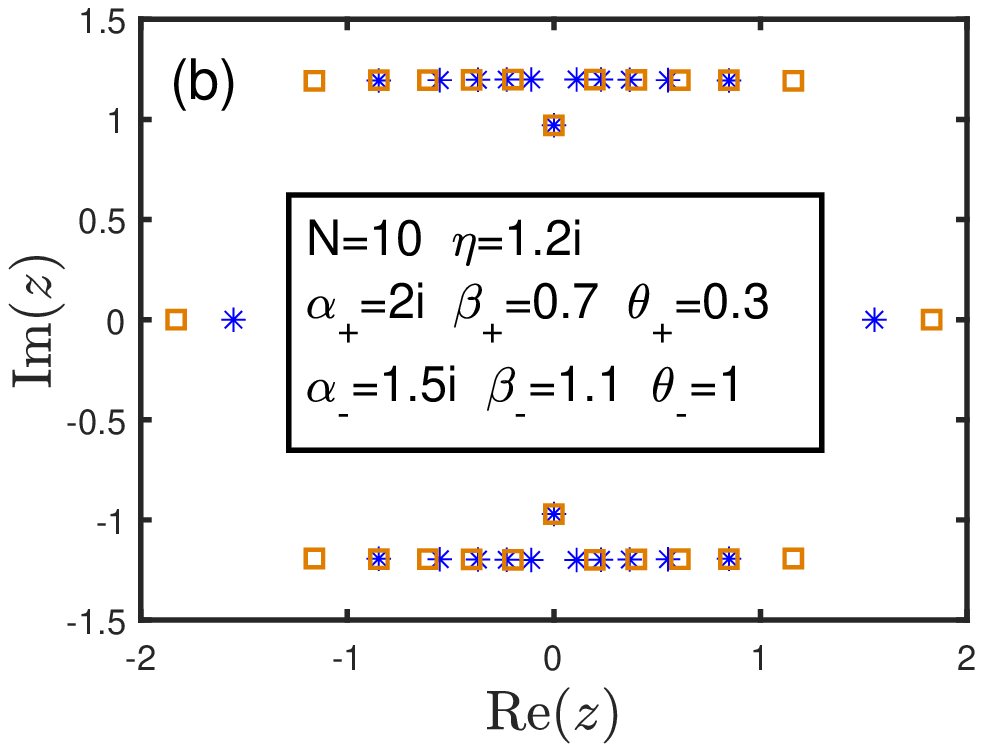}\vspace{-0.2cm}
\caption{Exact numerical diagonalization results of the root distributions in the ground state for $N=10$. (a)The blue asterisks indicate $\bar{z}$-roots for $\{\bar{\theta}_j=0\}$ and the brown squares specify $\bar{z}$-roots with the inhomogeneity parameters $\{\bar{\theta}_j=0.5j\}$. (b)The blue asterisks indicate $z$-roots for $\{\theta_j=0\}$ and the brown squares specify $z$-roots with the inhomogeneity parameters $\{\theta_j=0.1j\}$. }\label{Eg-theta}
\end{figure}

A plausible fact is that by choosing a proper set of inhomogeneity parameters, the root distributions possess manageable patterns in the thermodynamic limit. For example, for real $\eta$, we choose all $\{\theta_j\}\equiv \{i{\bar\theta}_j\}$ to be imaginary. As shown in Fig.\ref{Eg-theta}(a), the imaginary inhomogeneity parameters almost do not affect the imaginary parts of the roots $\bar{z}_{j}\equiv -iz_j$ but the distribution along the real axis, which allows us to derive the density of roots via Fourier transformation with an auxiliary function $\sigma(\bar{\theta})$, a given density of the inhomogeneity. The density of roots of the corresponding homogeneous system can then be obtained by finally taking the homogeneous limit $\sigma(\bar{\theta})\to\delta(\bar{\theta})$. For the imaginary $\eta$ case, as shown in Fig.\ref{Eg-theta}(b), we can follow the same procedure by using real inhomogeneity parameters.

Now let us turn to consider the surface energy defined by $E_{b}=E_{g}-E_p$, where $E_g$ is the ground state energy of the present system and $E_p$ is the ground state energy of the corresponding periodic chain \cite{book}. For a real $\eta>0$, there is a finite gap in the bulk spectrum. All the boundary parameters must be real to ensure a hermitian Hamiltonian (\ref{Hamiltonian}). For arbitrary imaginary inhomogeneity parameters, from the intrinsic properties of the $R$-matrix one can easily prove that $t^\dagger(u)=t(u^*)$ and $\Lambda^*(u)=\Lambda(u^*)$.
Due to the periodicity of $\Lambda(u)$, we fix the real parts of $\bar{z}_{j}$ in the interval $(-\frac\pi2, \frac\pi2]$. The roots can be classified into (i)real $\pm\bar{z}_j$; (ii)$\pm\bar{z}_j$, $\pm \bar{z}_j^*$ conjugate pairs with imaginary parts around $\pm\frac{in\eta}2$ ($n\geq 2$) \cite{Qia20} and (iii)conjugate pairs induced by the boundary fields either at the origin or at Re$\{{\bar z}_j\}=\frac\pi2$. We remark that the structure of the bulk conjugate pairs is quite similar to the string structure of the Bethe roots in the periodic chain \cite{tak} and the boundary conjugate pairs are tightly related to the boundary bound states appeared in the diagonal boundary case \cite{skorik}. In addition, one can easily prove that the energy is invariant under the parameter changes: (i)$\alpha_\pm\to-\alpha_\pm$, (ii)$\beta_\pm\to-\beta_\pm$, (iii)$\alpha_+\to-\alpha_+, \beta_+\to-\beta_+, \theta_+\to \pi+\theta_+$, (iv)$\alpha_-\to-\alpha_-, \beta_-\to-\beta_-, \theta_-\to\pi+\theta_-$ and (v) $\beta_+\to\beta_-, \beta_-\to\beta_+$.
Therefore, we consider only the case of $\alpha_\pm, \beta_+>0$ and $|\beta_+|\geq|\beta_-|$. It is sufficient to quantify the boundary contributions by tuning $\beta_{-}$ in four regimes: (I)$\beta_+>\beta_{-}>\eta/2$,
(II)$\eta/2\geq\beta_-\geq 0$, (III)$0>\beta_{-}>-\eta/2$ and (IV)$-\eta/2\geq\beta_{-}>-\infty$.

\begin{figure}[ht]
\centering
\includegraphics[width=4.2cm,height=3.2cm]{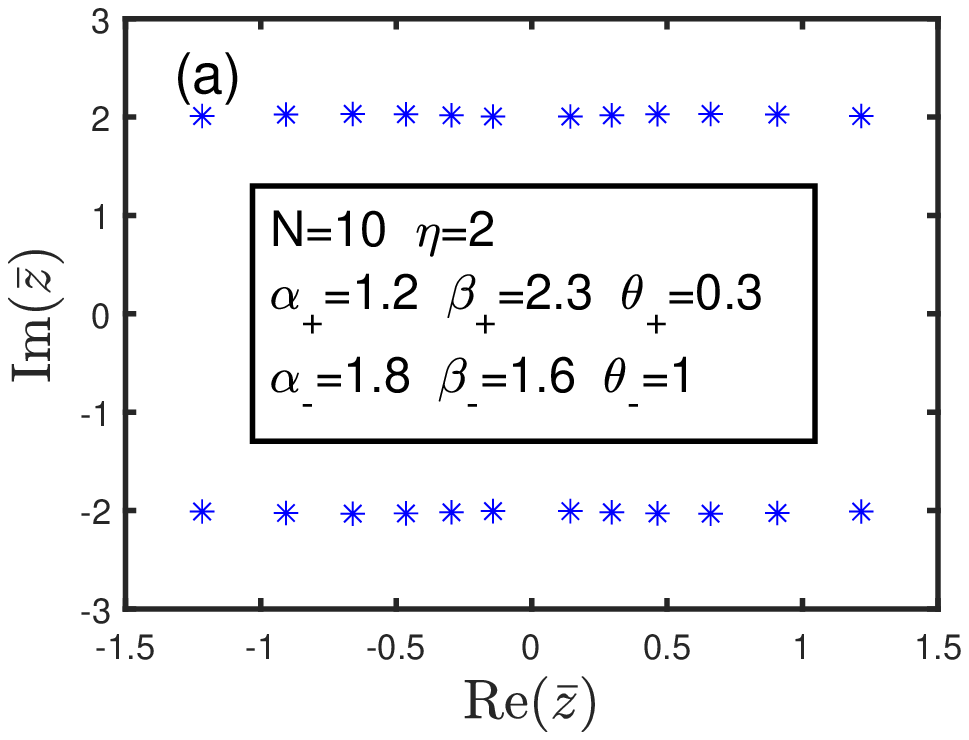}\hspace{0.1cm}
\includegraphics[width=4.2cm,height=3.2cm]{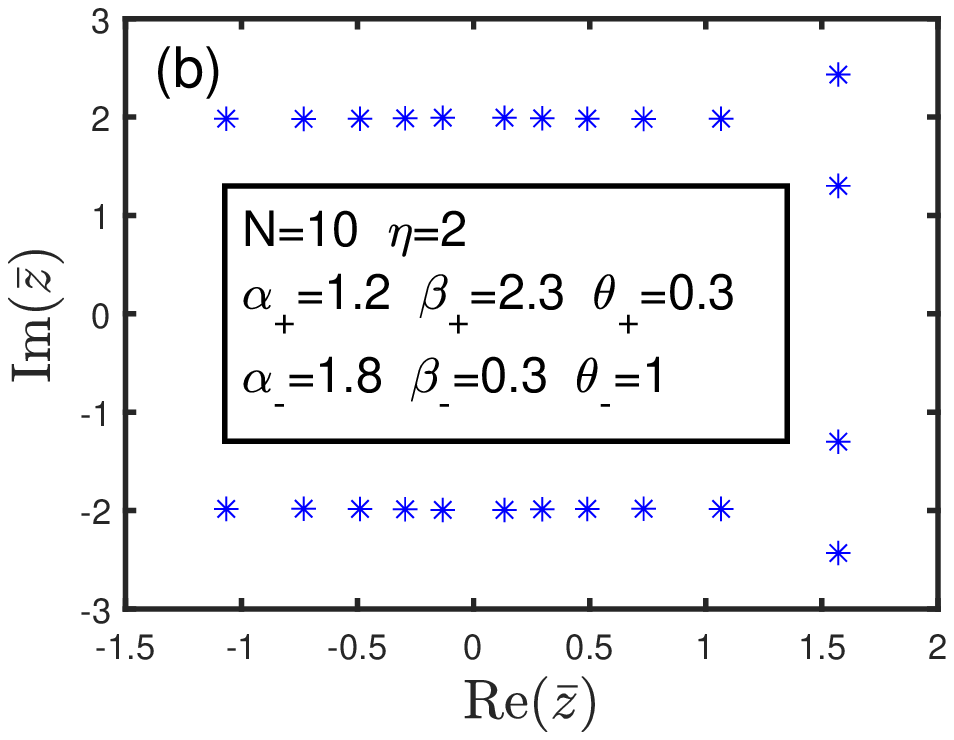}\\ \vspace{0.2cm}
\includegraphics[width=4.2cm,height=3.2cm]{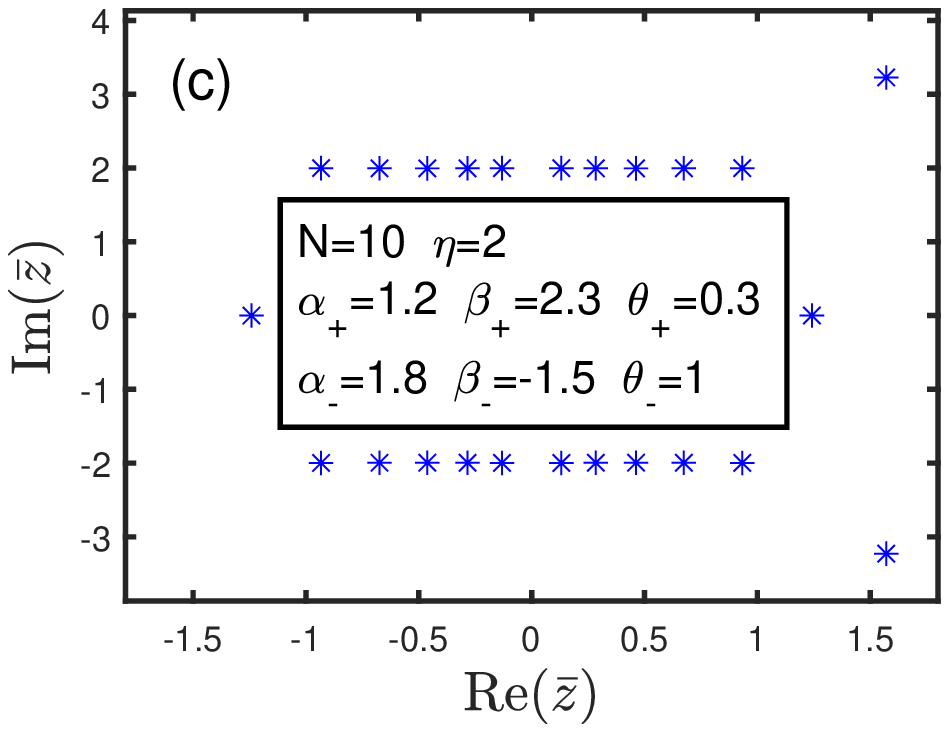}\hspace{0.1cm}
\includegraphics[width=4.2cm,height=3.2cm]{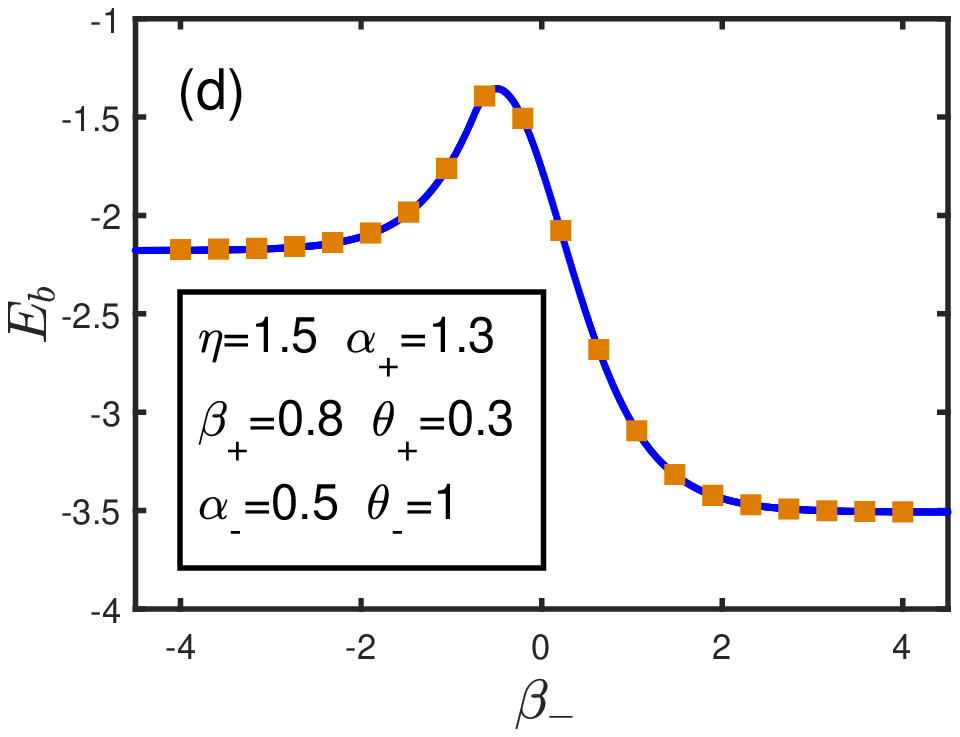} \vspace{-0.3cm}%,height=3cm
\caption{(a)-(c) Exact numerical results of $\bar z$-roots of the ground state in different regimes of boundary parameters for $N=10$ and $\eta=2$. (d) The surface energy versus $\beta_-$ for $\eta=1.5$. The blue line indicates the analytic result and the brown squares indicate the DMRG results for $N=212$.}\label{GS-real}
\end{figure}

We consider first the ground state for even $N$ and $\alpha_\pm, \beta_+>\eta/2$ case. In regime (I), all the $\bar z$-roots form conjugate pairs as $\{\bar{z}_j\sim\tilde{z}_j\pm\eta i| j=1,\cdots,N+2\}$ with real $\tilde{z}_j$. In the thermodynamic limit, the distribution of $\tilde z$-roots can be described by a density per site $\rho(\tilde{z})$. Furthermore, we assume that the inhomogeneity also has a continuum density per site $\sigma(\bar{\theta}_j)\sim 1/N(\bar{\theta}_j-\bar{\theta}_{j-1})$. Taking the logarithm of Eq.(\ref{tw1}) and making the difference of the equations for $\theta_j$ and $\theta_{j-1}$, by omitting the $O(N^{-1})$ terms we readily have
\bea
&&\hspace{-0.5truecm} N\int_{-\frac\pi2}^{\frac\pi2}[b_2(u-\bar{\theta})+b_2(u+\bar{\theta})] \sigma(\bar{\theta})d\bar{\theta}+ b_2(u-\frac\pi2) \no \\
&&\hspace{-0.5truecm} +b_2(u)+b_{\frac{2\beta_{-}}{\eta}}(u-\frac\pi2) +b_{\frac{2\beta_{+}}{\eta}}(u-\frac\pi2) +b_{\frac{2\alpha_{-}}{\eta}}(u) \no \\
&&\hspace{-0.5truecm} +b_{\frac{2\alpha_{+}}{\eta}}(u)= N\int_{-\frac\pi2}^{\frac\pi2}[b_1(u-\tilde{z})+b_3(u-\tilde{z})]\rho(\tilde{z})d\tilde{z} \no \\
&&\hspace{-0.5truecm} +b_1(u)+b_1(u-\frac\pi2),\label{ai}
\eea
where $b_n(x)  =\cot(x+\frac{n\eta i}{2})+\cot(x-\frac{n\eta i}{2})$.
The above equation is a convolution one and allows us to make Fourier transformation
\bea
&&\hspace{-0.5truecm}\tilde{\rho}(k)=[2N\tilde{b}_2\tilde{\sigma}(k)+[1+(-1)^k](\tilde{b}_2-\tilde{b}_1) +\tilde{b}_{\frac{2\alpha_{+}}\eta} \no \\
&&+\tilde{b}_{\frac{2\alpha_{-}}\eta} +(-1)^k(\tilde{b}_{\frac{2\beta_{+}}\eta} +\tilde{b}_{\frac{2\beta_{-}}\eta})]/[N(\tilde{b}_1+\tilde{b}_3)],\label{rho-r1}
\eea
where the Fourier spectrum $k$ takes integer values and $\tilde{b}_n(k)=-2 sign(k)\pi i e^{-\eta|nk|}$.
In the homogeneous limit, we take $\sigma(\bar{\theta})=\delta(\bar{\theta})$.
The ground state energy of the Hamiltonian (1) can thus be expressed as
\bea
E_{g1}=\frac{N i\sinh\eta}{2\pi}\sum_{k=-\infty}^{\infty}[\tilde{a}_1(k)- \tilde{a}_3(k)]\tilde{\rho}(k)-c_0,
\eea
where $\tilde{a}_n(k)=2\pi ie^{-\eta|nk|}$ is the Fourier transformation of
$a_n(x)=\cot(x-\frac{n\eta i}{2})-\cot(x+\frac{n\eta i}{2})$.
We note that the boundary parameters $\theta_{\pm}$ do not appear in (\ref{ai}), implying that they contribute nothing to the surface energy in the leading order. Direct calculation gives the surface energy $E_{b1}$ in regime (I) as
\bea
&&E_{b1}=e_b(\alpha_+,\beta_+)+e_b(\alpha_-,\beta_-)+e_{b0},\no\\
&&e_b(\alpha,\beta)=-2\sinh\eta\sum_{k=1}^{\infty}\tanh(k\eta)\{(-1)^ke^{-2k\eta}\no\\
&&\qquad \qquad+e^{-2k|\alpha|}+(-1)^ke^{-2k|\beta|}\}-\tanh\eta\sinh\eta,\no\\
&&e_{b0}=-2\sinh\eta\sum_{k=1}^{\infty}\{\tanh(k\eta)[1-(-1)^k]e^{-2k\eta}\no\\
&&\qquad\quad -[1+(-1)^k]e^{-k\eta}\}+\tanh\eta\sinh\eta,\label{Eb1}
\eea
where $e_b(\alpha,\beta)$ indicates the contribution of one boundary field and $e_{b0}$ is the surface energy induced by the free open boundary \cite{skorik}.

In regime (II), besides the bulk conjugate pairs around the $\pm\eta i$ lines, there exist two boundary conjugate pairs  $\frac\pi2\pm(\beta_{-}+\frac\eta2)i$ and
$\frac\pi2\pm(\beta_{+}+\frac\eta2)i$ fixed by (\ref{tw2}) as shown in Fig.\ref{GS-real}(b).
Taking the boundary roots into account, with a similar procedure used in regime (I) we find that the bare contribution of the boundary conjugate pairs to the energy is exactly cancelled by that of the back flow of the continuous root density, as happened in the diagonal boundary case \cite{skorik}. The surface energy $E_{b2}$ takes exactly the same form of (\ref{Eb1}). Taking $\alpha_-\to\infty$ and $\beta_-\to 0$, $\vec{h}_-=0$, $e_b(\infty,0)=0$.
Therefore, the contributions of the two boundary fields and the free open boundary to the surface energy are additive in regimes (I) and (II).

In regime (III), there also exist two boundary conjugate pairs. However, the absolute value of the imaginary part of the inner conjugate pair is $\beta_-+\frac\eta2<\frac\eta2$. In this case, the inner boundary conjugate pair indeed contributes a nonzero value to energy and the surface energy reads
\bea
&&E_{b3}=4\sinh\eta\sum_{k=1}^{\infty}(-1)^k e^{-k\eta}\tanh(k\eta)\cosh(2k\beta_-\!+\!k\eta)\no \\
&&\quad +E_{b1}+\sinh\eta[\tanh(\beta_{-}+\eta)-\tanh(\beta_{-})].\label{Eb2-2}
\eea
The contributions of the two boundaries to the surface energy are no longer additive and a correlation effect between the two boundary fields appears.

In regime (IV), only one boundary conjugate pair exist as shown in Fig.\ref{GS-real}(c). However, due to the symmetry of root distribution, two real roots around $\pm \frac \pi 2$ must exist. The boundary conjugate pair contributes nothing to the surface energy but the two real roots do contribute a nonzero value to energy and the surface energy reads
\bea
&&\hspace{-0.5truecm}E_{b4}=E_{b1}+E_h,\no \\
&&\hspace{-0.5truecm}E_h=2\sinh\eta\Big[\sum_{k=1}^{\infty}\frac{(-1)^k2\tanh(k\eta)}{ e^{k\eta}}+\tanh\frac\eta2\Big].\label{Eb3}
\eea
In this regime, the correlation effect of the two boundaries does not rely on the magnitudes of the boundary fields but on the sign of $\beta_+\beta_-$.

\begin{figure}[ht]
\centering
\includegraphics[width=4.2cm,height=3.3cm]{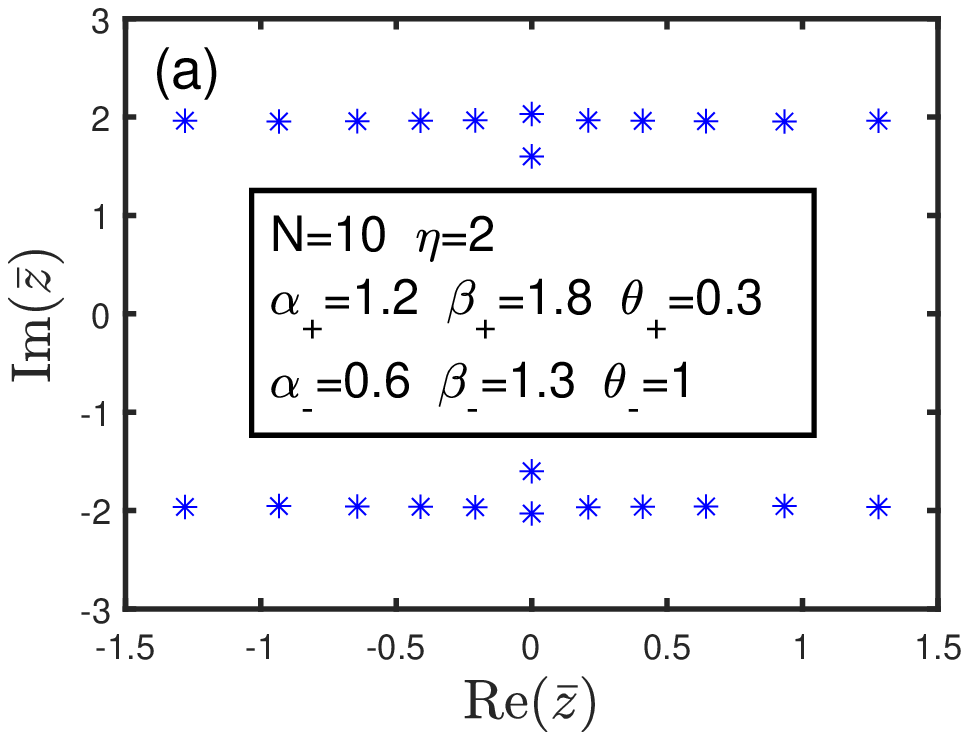}\hspace{0.1cm}
\includegraphics[width=4.2cm,height=3.3cm]{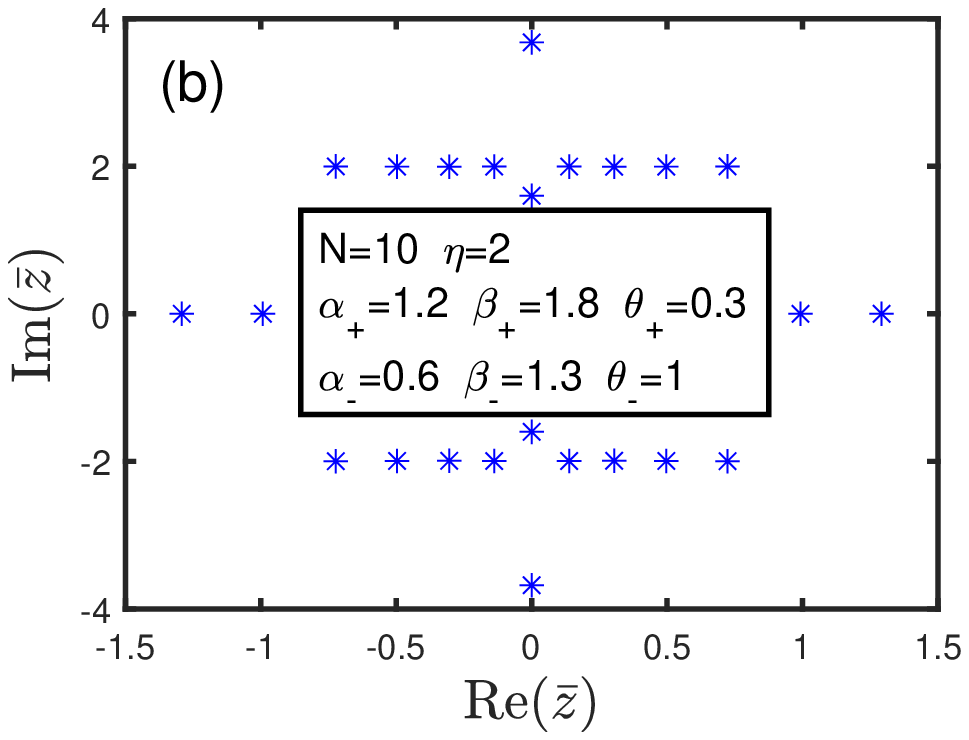}\vspace{-0.2cm}
\caption{The distribution of $\bar z$-roots for $N=10$ and $\eta=2$. (a) The ground state. (b) A low-lying excited state.}\label{Ee-real}
\end{figure}

We note that if $|\beta_\pm|<\eta/2$ and $\beta_+\beta_-<0$, we can always choose $\beta_-$ as $-\min\{|\beta_+|,|\beta_-|\}$ and $\beta_+$ as $\max\{|\beta_+|,|\beta_-|\}$ in (\ref{Eb2-2}) to get the correct surface energy. For comparison, the density matrix renormalization group (DMRG) method \cite{white} is performed for $N=212$ and several values of $\beta_-$ . Our analytic results coincide perfectly with the numerical ones as shown in Fig.\ref{GS-real}(d). For $\alpha_{\pm}\in(0,\eta/2)$, central conjugate pairs associated with the boundary fields around $\pm i(\eta/2+\alpha_{\pm})$ exist in the ground state as shown in Fig.\ref{Ee-real}(a). Exact calculation shows that these boundary roots contribute nothing to the surface energy in the thermodynamic limit, as their contributions are exactly cancelled by that of the back flow of the bulk root density. By examining the root patterns we obtain that the surface energy $E_{b}^{odd}(\beta_-)$ for an odd $N$ can be given by $E_{bl}(-\beta_-)$ for an even $N$ as $E_b^{odd}(\beta_-)=E_{bl}(-\beta_-)-E_h$, where $l=1,2,3,4$ indicates the corresponding boundary parameter regime. Such a parity dependence of the surface energy is in fact due to the long range Neel order in the bulk. For an even $N$ the two boundary spins prefer to be anti-parallel, while for an odd $N$ the two boundary spins prefer to be parallel. Therefore, fixed boundary fields must induce different surface energies for even $N$ and odd $N$ in the thermodynamic limit $N\to\infty$.

Usually, exact fractional excitations can be derived in most of the integrable models with $U(1)$ symmetry. A typical kind of fractional excitations in the periodic spin chain model is spinon, which is believed carrying spin-$\frac12$ \cite{tak-fad}. In the open boundary case, the unparallel boundary fields break the $U(1)$ symmetry and the $z$-component of the total spin is no longer a good quantum number. A question thus arises: Is there any spinon-like excitations in the present system? As an example to answer this question, let us consider a simple root distribution away from that of the ground state in regime (I): taking two conjugate pairs away in the ground state configuration
and adding four real roots on the real axis. The four real roots are distributed symmetrically around the origin as required by the symmetry of the eigenvalue function $\Lambda(u)$. In addition, two imaginary conjugate pairs may appear in the root configuration
as shown in Fig.\ref{Ee-real}(b). We denote the four real roots as $\pm z_1$ and $\pm z_2$. The excitation energy in the thermodynamic limit associated with this root pattern can be derived by following the same procedure discussed in the previous text
\bea
E_e=\varepsilon(z_1)+\varepsilon(z_2), \varepsilon(z)=2\sinh\eta\sum_{k=-\infty}^{\infty}\frac{e^{-2ikz}} {\cosh(k\eta)}.
\eea
It seems that the excitation energy only depends on the positions of the real roots and takes exactly the same dispersion form of spinons in the periodic chain. Even though, such kind of elementary excitations should be rather different from the traditional spinons \cite{tak-fad} due to the broken $U(1)$-symmetry. In fact, these excitations must be helical in the real space to match the two unparallel boundaries. The helical structure can be characterized either by the quantity $\langle \vec \sigma_j\times \vec \sigma_{j+1}\rangle$, which is nonzero in the non-diagonal boundary cases but zero in the parallel boundary cases, or by the structure of the eigenvectors constructed from a helical pseudo-vacuum state\cite{cao03, Zhan15}.

\begin{figure}[ht]
\centering
\includegraphics[width=4.2cm,height=3.2cm]{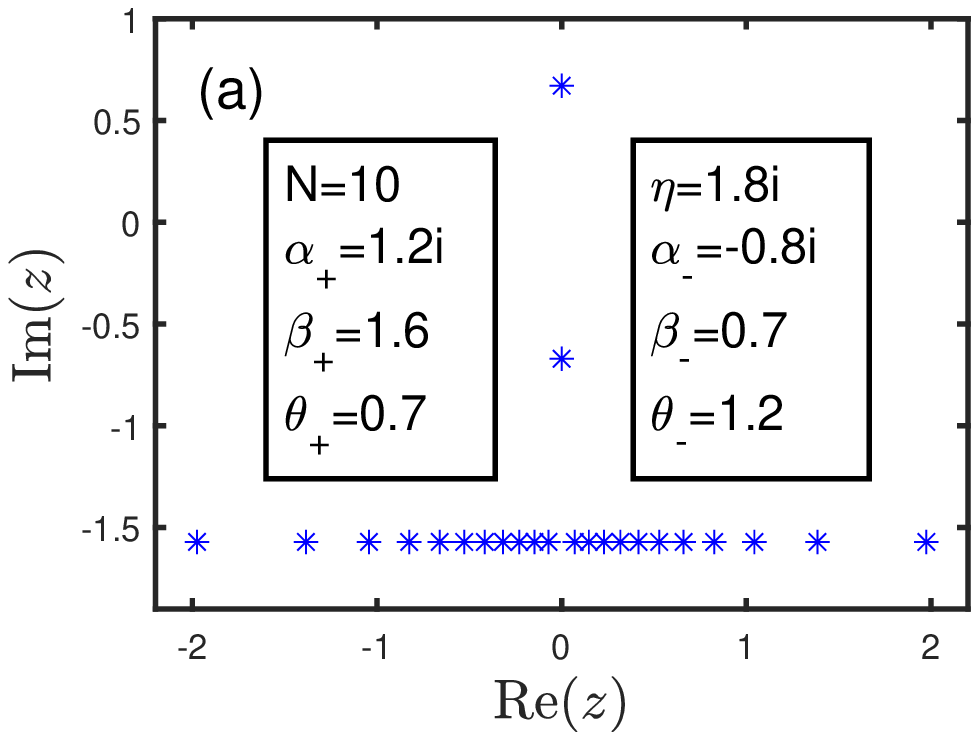}\hspace{0.1cm}
\includegraphics[width=4.2cm,height=3.2cm]{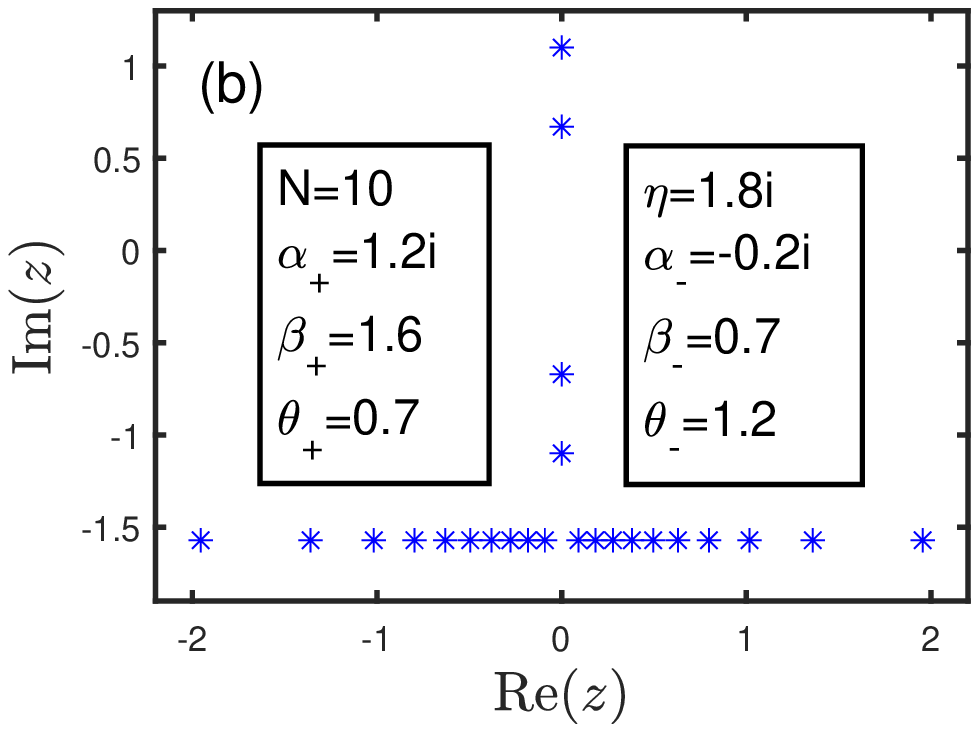}\\ \vspace{0.2cm}
\includegraphics[width=4.2cm,height=3.2cm]{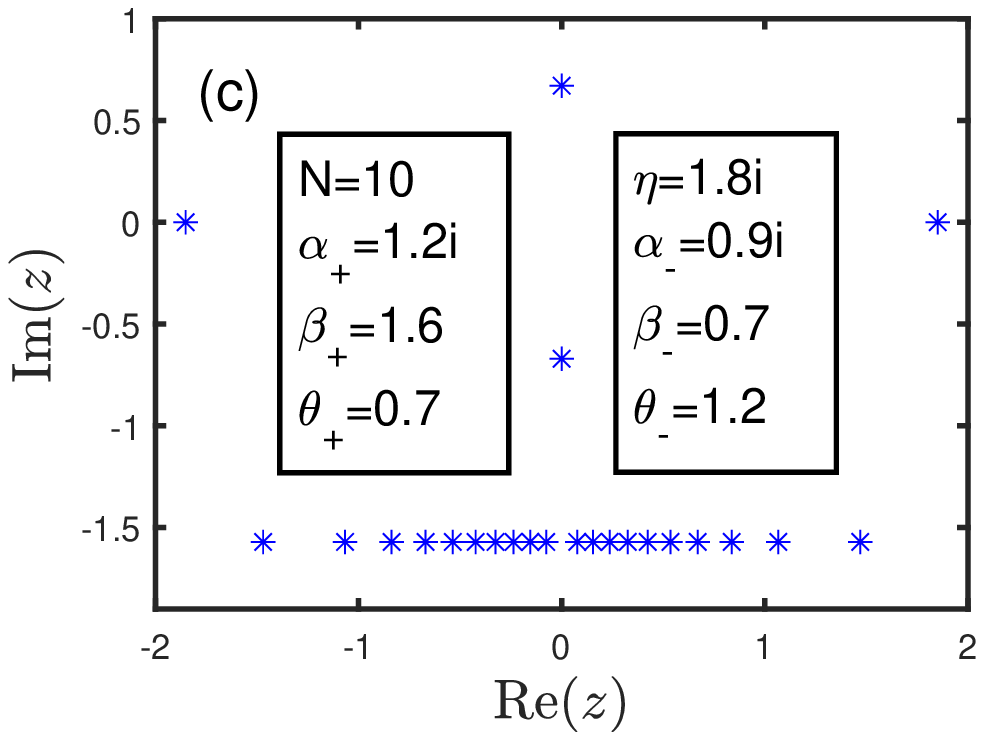}\hspace{0.1cm}
\includegraphics[width=4.1cm,height=3.2cm]{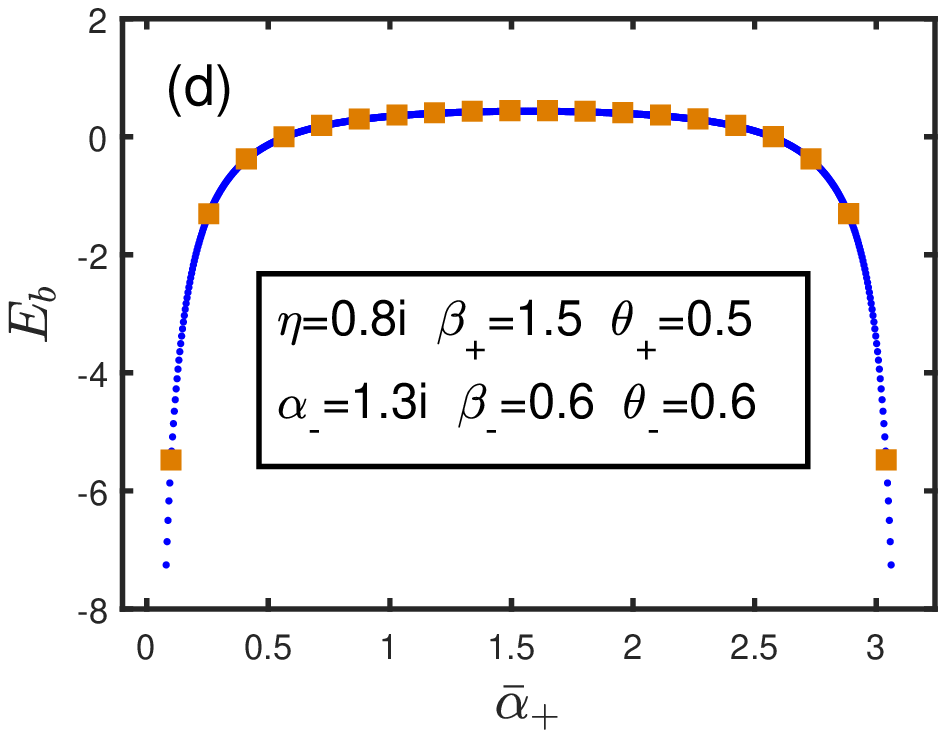}\vspace{-0.2cm}
\caption{(a)-(c) $z$-roots of the ground state for $N=10$, $\eta=1.8i$ and  given sets of boundary parameters. (d) The surface energy versus the boundary parameter $\bar{\alpha}_+$ for $\eta=0.8i$.
The blue dots indicate the analytic results and the brown squares are the ones obtained via DMRG for $N=212$.}\label{GS-imag}
\end{figure}

For an imaginary $\eta$, the spectrum of the Hamiltonian (1) is massless. We take imaginary $\alpha_\pm$ and real $\beta_\pm$ and $\theta_\pm$ to define real boundary fields. By choosing real inhomogeneity parameters, we have $t^\dagger(u)=t(-u^*)$ and $\Lambda^*(u)=\Lambda(-u^*)$. The roots can be classified into (i)real $\pm z_j$; (ii)on the line Im$\{z_j\}=-\frac\pi2$; (iii)bulk conjugate pairs Im$\{z_j\}\sim\pm\frac{in\eta}2$ ($n\geq 2$) and (iv)central conjugate pairs associated with the boundaries.
For convenience, let us introduce the notations $\gamma=-i\eta$ with $\gamma\in(0,\pi)$ and $\bar{\alpha}_{\pm}=-i\alpha_{\pm}$. Without losing generality, we restrict $-\frac\pi2\leq$ Im$\{z_j\}<\frac \pi2$ for the periodicity of $\Lambda(u)$.

For $\gamma\in[\frac\pi2, \pi)$, the $z$-roots in the ground state for a given set of boundary parameters and $N=10$ is shown in Fig.\ref{GS-imag}(a). Most of the roots locate on the line $-i\frac\pi 2$ and one conjugate  pair $\pm \frac{i(\pi-\gamma)}2$ locates on the imaginary axis. The existence of this conjugate pair does not depend on the values of the boundary parameters. By tuning the value of $\beta_-$, we find that the structure of $z$-roots keeps unchanged, which indicates that the ground state energy is given by an unified formula for arbitrary real boundary parameters $\beta_\pm$.

By varying $\alpha_-$,
a central conjugate pair $\pm i(\frac\gamma2+|\bar{\alpha}_-|)$ appears when $|\bar{\alpha}_-|\in(0,\frac{\pi-\gamma}2)$ as shown in Fig.\ref{GS-imag}(b). Direct calculation shows that the central conjugate pairs do not contribute to the surface energy. The above conclusion also holds for $\bar{\alpha}_+$. Besides, depending on $\bar{\alpha}_{\pm}$ and the parity of $N$, two real roots may exist at the boundaries as shown in Fig.\ref{GS-imag}(c). These roots tend to $\pm\infty$ in the thermodynamic limit and also do not contribute to the surface energy.
For the case corresponding to Fig.\ref{GS-imag}(a), in the thermodynamic limit the density of roots satisfies
\bea
&&\hspace{-0.5truecm}N\int_{-\infty}^{\infty} [b_2(u-\theta)+b_2(u+\theta)]\sigma(\theta)d\theta +b_2(u) \no \\ &&\hspace{-0.5truecm}
+\frac12[b_{\frac{\pi}{\gamma}}(u+\beta_+)+b_{\frac{\pi}{\gamma}}(u-\beta_+) +b_{\frac{\pi}{\gamma}}(u+\beta_-)\no \\ &&\hspace{-0.5truecm}
+b_{\frac{\pi}{\gamma}}(u-\beta_-)] +b_{\frac{2\bar{\alpha}_{+}}{\gamma}}(u)+b_{\frac{2\bar{\alpha}_{-}}{\gamma}}(u) = b_{\frac\pi\gamma-1}(u) \no
\\ &&\hspace{-0.5truecm} +b_{\frac\pi\gamma}(u)+b_1(u)+ N\int_{-\infty}^{\infty}b_{\frac\pi\gamma-1}(u-z)\rho(z)dz,
\eea
where $b_n(x)={\rm csch}^2(x+\frac{n\gamma i}{2})+{\rm csch}^2(x-\frac{n\gamma i}{2})$. Taking the Fourier transformation and homogeneous limit
$\sigma(\theta)\to\delta(\theta)$,
we finally obtain the surface energy
\bea
&&\hspace{-0.5truecm}E_b=-\frac{\sin\gamma}2\int_{-\infty}^{\infty}\frac{\tanh(\frac{k\gamma}2)}{\sinh(\frac{k\pi}2)} \{\cosh\frac{k(\pi-2\gamma)}2-1 \no \\
&& \quad +\cosh\frac{k(\pi-2\bar{\alpha}_+ +2\pi\lfloor \frac{\bar{\alpha}_+}\pi )\rfloor)}2 +\cos\beta_+\no \\
&&\quad +\cosh\frac{k(\pi-2\bar{\alpha}_-+2\pi\lfloor \frac{\bar{\alpha}_-}\pi )\rfloor)}2 +\cos\beta_- \no \\
&&\quad -\cosh\frac{k\gamma}2 -\cosh\frac{k(\pi-\gamma)}2\}dk.\label{Eb-Imag}
\eea

For $\gamma\in(0, \frac\pi2)$, most of the roots in the ground state locate on the lines $\pm i\gamma$ and the rest roots form central conjugate pairs as shown in Fig.\ref{Eg-theta}(b). The surface energy is still given by Eq.\eqref{Eb-Imag}. Comparison of the DMRG results and our analytic results is given in Fig.\ref{GS-imag}(d). The present result also coincides exactly with that derived in \cite{liyuanyuan}. The absence of correlation and parity effects is due to the absence of long-range order in the gapless bulk.

In conclusion, an analytic method is developed to obtain the surface energy and elementary excitations of the XXZ spin chain with generic non-diagonal boundary fields in both gapped and gapless regimes. This method provides an universal procedure to compute physical quantities of quantum integrable systems either with or without $U(1)$ symmetry \cite{book,npb2,hep4} in thermodynamic limit.

The financial supports from  MOST (2016YFA0300600 and
2016YFA0302104),
NSFC (12074410, 12047502, 119 34015,
11975183, 11947301 and 11774397), the Strategic
Priority Research Program of CAS (XDB33000000)
and the fellowship of China Postdoctoral Science Foundation (2020M680724) are gratefully acknowledged.

\end{document}